\documentclass[runningheads,a4paper]{llncs}

\usepackage{amssymb}

\usepackage{graphicx}
\usepackage{algorithm}
\usepackage{algorithmic}
\usepackage{url}

\urldef{\mailsa}\path|kkhingwe@gmail.com,msb@nitt.edu|    
\newcommand{\keywords}[1]{\par\addvspace\baselineskip
\noindent\keywordname\enspace\ignorespaces#1}

\begin{document}
\pagestyle{empty} 
\mainmatter  

\title{Hierarchical Role-Based Access Control with Homomorphic Encryption for Database as a Service}

\titlerunning{Hierarchical RBAC with Homomorphic Encryption for DBaaS}

%
%
\author{Kamlesh Kumar Hingwe%
\and S. Mary Saira Bhanu}
\authorrunning{Hierarchical RBAC with Homomorphic Encryption for DBaaS}

\institute{Department of Computer Science and Engineering\\National Institute of Technology\\
Tiruchirappalli, India\\
\mailsa}

%
%

\toctitle{Lecture Notes in Computer Science}
\tocauthor{Authors' Instructions}
\maketitle

\begin{abstract}
Database-as-a-service provides services for accessing and managing customer’s data which provides ease of access, and the cost is less for these services. There is a possibility that the DBaaS service provider may not be trusted, and data may be stored on a untrusted server. The access control mechanism can restrict users from unauthorized access, but in cloud environment access control policies are more flexible. However, an attacker can gather sensitive information for a malicious purpose by abusing the privileges as another user and so database security is compromised. The other problems associated with the DBaaS is to manage role hierarchy and secure session management for query transaction in the database. In this paper, a role based access control for the multi-tenant database with role hierarchy is proposed. The query is granted with least access privileges, and a session key is used for session management. The proposed work protects data from privilege escalation and SQL injection. It uses the partial homomorphic encryption (Paillier Encryption) for the encrypting the sensitive data. If a query is to perform any operation on sensitive data, then extra permissions are required for accessing sensitive data.  Data confidentiality and integrity are achieved by using the role-based access control with partial homomorphic encryption.

\end{abstract}

\section{Introduction}

The Databases used in the cloud can be of NoSQL type (Amazon, SimpleDB, Yahoo PNUT, CouchDB) or SQL type (Oracle and MySQL). In cloud, a service called Database as a Service (DBaaS) is provided for database installation and maintenance. The service provides features such as on-demand independent service for managing the data, instant access to the ubiquitous service, etc.
\par The DBaaS provides the cost reduction, easy database maintenance, performance tuning and it also supports multi-tenancy. The use of the multi-tenant database in cloud computing leads to security challenges, due to resource sharing \cite{takabi2010security}. The cloud users compromise the security of their data and computing application for cheaper service because the data management for the large number of users is not easy, and the database configurations are creating the vulnerabilities.
\par The DBaaS security issue is due to the absence of user authentication mechanism, data authorization, lack of session management and unsecured key management system. The DBaaS threats \cite{bertino2011access} are as follows:
\begin{enumerate}
  \item \textbf{Confidentiality Threats: -} The information is stored on the database server. The authorized user should get the access to the database. If data access is through DBaaS and if encryption mechanism is weak then data will be compromised with confidentiality. The attacks related to confidentiality are
  \begin{enumerate}
    \item \textbf{Insider attack :-} The Database administrator has all the privileges to access the database for the maintenance purpose. If the privilege is misused, then it leads to big confidentiality threat.
    \item \textbf{Outsider Attack :-} The outsider attacks in DBaaS are related to the exploitation of the software vulnerability. The attacker can also do the spoofing, side channeling, and man in middle attacks. Intrusion is also a problem of the DBaaS in which the attacker access the login credentials.
    \item \textbf{Access Control Issues :-} Policies of access control are managed by the DBaaS provider not by the data owner. So due to lack of the monitoring system, the access control policies cannot be customized. 
  \end{enumerate}
  \item \textbf{Integrity Threats :-} The data integrity refers to protecting the data from unauthorized modification or deletion. The data tempering in DBaaS can happen at any level of storage.
\end{enumerate}
The access control mechanism plays an important role to provide DBaaS security such as data confidentiality and integrity \cite{bertino2011access}. It validates the rights of users against the set of the authorization rules and states to perform the operation on the database. The authorization states are dependent on the access control policies of the database owner or organization. The confidentiality of the database is improved by encrypting the database and the data integrity is provided by the access control mechanism. The user's database access requests are properly authorized by enforcing the access control polices.
\par The access control mechanisms commonly used are Mandatory Access Control (MAC), Discretionary access control (DAC) \cite{bertino2011access} and Role-based access control (RBAC). The RBAC is used for the traditional databases and it assigns permissions according to the role of the user in the organization. The three primary rules of RBAC are role assignment, role authorization, and permission authorization. RBAC is followed in this paper.
\par The use of access control provides security to DBaaS but still there are some challenges \cite{bertino2005database} such as Integration of DBaaS access control with efficient role hierarchy management, and session management. This paper proposes an access control mechanism that performs the user authentication using roles, and group authentication using group key. The proposed work also maintains secure session for the query transactions, maintain the role hierarchy and secures data using Paillier encryption.
 
\section{Related Work}
Zhe Jia et al. \cite{jia2012privacy} proposed a privacy-preserving access control that is based on secret sharing and ElGamal homomorphic properties. It provides security to the data owner, but also provides protection for the data requester using the access control policies.  Their model did not considered the multi-tenant database architecture.
\par Calero et al. \cite{calero2010toward} proposed a model for authorization that supports multi-tenancy role-based access control. The path back object hierarchy is used. Hu et al. \cite{hu2009towards} proposed a work that is based on distributed role base control for web technology. Echeveria et al.\cite{echeverria2010permission} proposed a solution for the de-coupling access control using attribute encryption.
\par The Siebel systems \cite{brodersen2004database} used a role based access control mechanism that allows multiple tenants, in which each tenant is the owner of his separate virtual database. This mechanism supports an access control subsystem for multi user database access, where the each user has access of at least one organizational database attribute.  They divide the database into files, files into the records, and record into the fields. This mechanism is based on separation of individual database files, which is based on attribute ownership or granted access control. 

\par Yaish et al. \cite{yaish2011elastic,yaish2013multi} have proposed a multi-tenant database access control mechanism, called Elastic Extension Table Access Control (EETAC) that allows each tenant's with different type of access grants to access data in multi-tenant database. The mechanism of data retrieving in multi-tenant database is different from single-tenant database. The data of user's is subdivided by partitioning between particular tenants data and by access rows and columns granted to users, which is based on group and role assignment. The EETAC have role back access control for multi-tenant database but does not maintain the role hierarchy in the system.


\par Wu et al.\cite{wu2013acaas} proposed Access Control as a Service (ACaaS) for providing supports for the multiple access control model. They implemented role based access control which is configured for the amazon web services. The role hierarchy is maintained within the access control mechanism. The drawbacks of this method are it is designed only for amazon web services, and also it does not support multi-level access control policies and efficient session management for the user.

\par The existing works do not provide all solution at one place like secure session management, maintenance of role hierarchy, multi-level access control and data encryption. Some of existing \cite{jia2012privacy,wu2013acaas} work does not support the multi-tenant database architecture.

\section{Proposed work}
\begin{figure}
    \centering
    \includegraphics[width=90mm,keepaspectratio]{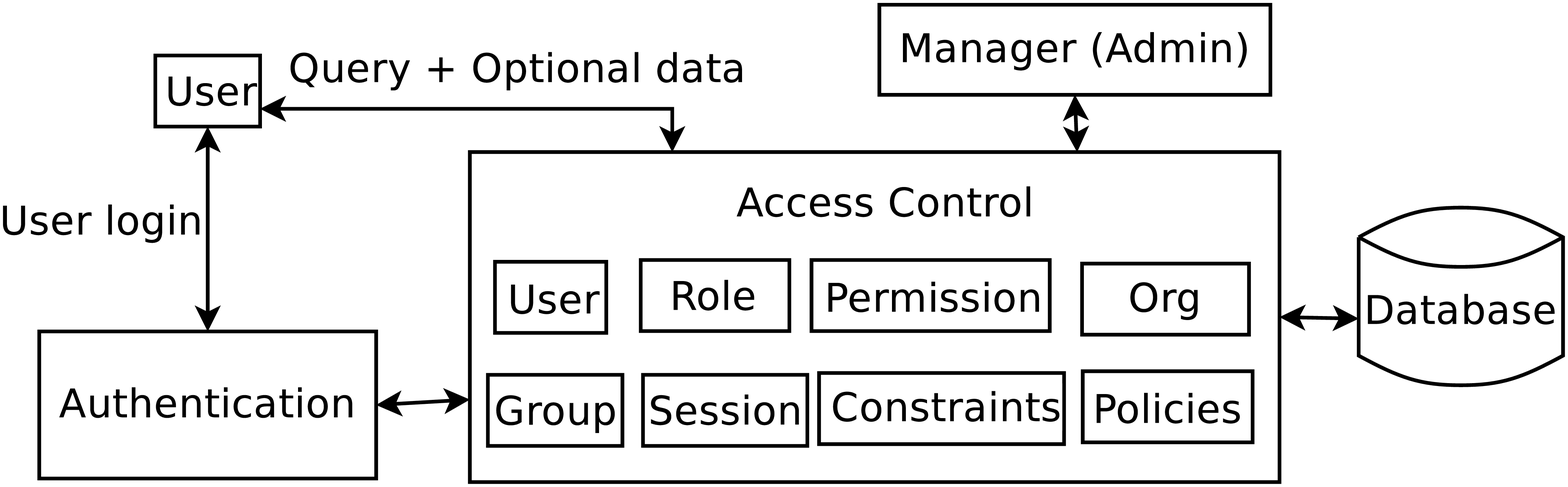}
    \caption{Block Diagram of Architecture}
\end{figure}
The proposed work uses role-based access control mechanism that will perform on the user's query. The proposed access control mechanism provides the confidentiality as well as integrity. The shared database and shared schema type isolation are used for the multi-tenant database.
\par The proposed work has two main modules; i) authentication and ii) access control. The authentication module is responsible for the user authentication and generation of the session key. It validates user's credentials by checking the user name and its password. After validating, a session key is issued to the user. This key is used for the data encryption that provides the secure communication between user and database owner. Session key is generated using the Secure Hash Algorithm (SHA-2).  The input for SHA-2 is a pseudo-random number, which is generated by the most efficient random number generator called Mersenne Twister \cite{matsumoto1998mersenne}. It generates integer in the range between 0 and $2^k$ -2 for k-bit word length. Queries are validated by the access control module and are redirected to the database for further processing. The access control module takes encrypted query, group key (if any group is assigned) and optional requests from users as well as it takes user id and session key as input from authentication module.
The administrator is responsible for maintenance of the role hierarchy. 
\begin{figure*}
    \centering
    \includegraphics[width=120mm,height=88mm]{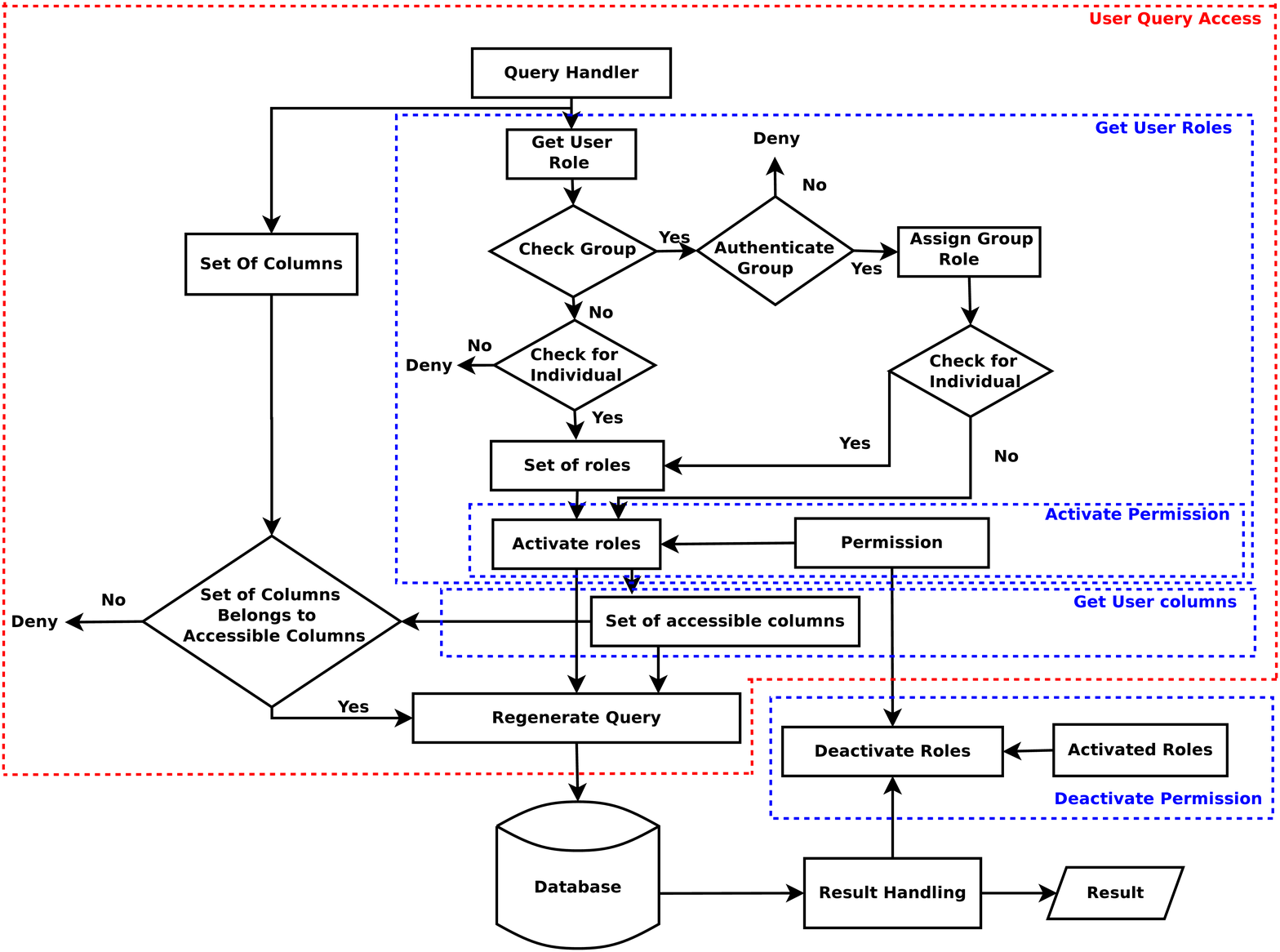}
    \caption{Flow Diagram of Access control Module}
\end{figure*}
\par The access control mechanism contains eight essential sub-modules as shown in fig.1. The modules are
\begin{enumerate}
  \item Org :- It is responsible for registration, deletion and listing of users in the DBaaS within a tenant.
  \item User :- It is responsible for managing all information of the user.
  \item Group :- It is responsible for managing the group in a single organization. An organization can have many groups. Moreover, each group have its own set of roles.
  \item Role :- Role module is responsible for creating and deleting the roles. It manages the role information in Database and also responsible for activation and deactivation of the role.
  \item Permission :- It contains all the set of rules for a role. The permission can be modified by Admin.
  \item Session :- It is responsible for maintaining the session for a complete query transaction.
  \item Constraints :- It contains set of rules to restrict some operations.
  \item Policies :- It contains a set of rules for providing access to the user's query.
\end{enumerate}

\par The proposed work provides data security for the tenant's and their customers by deploying the access control model in DBaaS. The authentication module at the server side generates a session key which is used by the client to encrypt the query. The access control module at the server decrypts the query using the same session key. The data sent by the client can be sensitive as well as non-sensitive. To ensure security, sensitive data is encrypted using a private key generated by server. The server processes the query and the result is sent to the client in the encrypted form. The result of the query is decrypted using the server public key at client machine.

\par The user's query needs right privileges for accessing database. For providing the secured resource access, a role has to be allotted to the user. These roles are binded with the permissions and constraints. If the user is part of a group, then a group key is also provided for accessing databases.

\par Fig. 2 represents the data flow diagram of access control module. The Query Handler takes user id, query, group key and optional requests as a single string of input; it parses users input and separates various attributes from the user input. First the role of the user is checked. The user may be allotted individual roles, or group roles or both. Once the role is identified, the assigned role is verified with the policies and constraints and the information is added in a set of roles. If user belongs to any group then perform group authentication using group key. After the successful authentication, add the group roles in the set of roles. The set of roles is given input to activate role for the role activation, which is responsible for the activating the roles by adding the permissions to access the resource. If the set of roles has more than one role, then this module search for the parent role in the given set and activate the parent role. However, if the parent role is not present in the given set, then add permissions for the role activation. After the role activation, get the set of columns in the query that are accessible to the user using the activated roles. If all queried columns are belonged to the accessible set of columns, then regenerate query according to the policies and constraints. The regenerated query takes input as the set of the accessible column for the user and the set of columns in the query. Then encrypt the sensitive data in the query using a Paillier encryption algorithm. The regenerated query is executed on the database, and the result of the query is sent to the result handler which is responsible for checking whether the query execution is successful. If successful, the result is sent to the user, otherwise if any error occurs, generate an error message. In both of the cases after the query execution, it triggers the Deactivate roles module which deactivate the user's role that are activated by the activate role module.
\par The access control module performs the user query access function which is responsible for database query execution and encrypted query regeneration. The three functions that are performing user query access function are:
\begin{enumerate}
  \item Get user roles :- It adds the assigned individual or group roles in the set of roles after the verification and validation of the roles. 
	\item Activate permission :- It is part of the get user role. It activates appropriate role to provide least access for the query from the set of roles.
	\item Get user column :- Get the accessible columns from the activated roles.
\end{enumerate}
\par The tables that are used to provide database access control are as follows:
\begin{enumerate}
\item Tenant Table :- It stores the information about the tenants.
\item User Table :- It can be more than one for a tenant in the database. 
\item Group Table :- It contains information or groups and associated user for the tenant. 
\item Role table :- It contains information about the roles. 
\item Permission Table :- It is used for the permission to the user that he is capable of performing a particular operation or not. 
\end{enumerate}
The access to the table depends on the access control policies and constraints because every table has their own access policies and constraints. 

\subsection{User Query Access Function}

The sensitive data of the query is encrypted using Paillier encryption algorithm \cite{gentry2009fully}. $QA_{ret}$ is obtained by passing all the set of policies and constraints associated with the users, group and role that is related to the requested query. The encryption is performed using private key for sensitive data in the query. The sensitive columns in the database are defined by the database owner. To access the sensitive data extra permission is required to users all the time for all roles (except admin role). The users have the public key that is known to the user that belongs to the role (role have permission to access sensitive data). The decryption takes place at user machine using the public key.
\par The key generation steps for the Paillier encryption are as follows:
\begin{enumerate}
  \item Select two independent large prime numbers $P_{1}$ and $P_2$ randomly such that $gcd(P_1P_2,(P_1-1)(P_2-1)=1)$ (if prime numbers of equal length)
  \item Compute  $n=P_1P_2$  and  $\gamma=lcm(P_1-1,P_2-1)$
  \item A random integer g is generated where $g\in \mathbb{Z}_{n}^{*}$
  \item Ensure \textit{n} divides the order of \textit{g} by checking the existence of the following modular multiplicative inverse  $\mu =(T(g^{\gamma}mod n^{2}))^{-1}mod n$  where \textit{T} is defined as  $T(u)=\frac{u-1}{n}$
  \item The private (encryption) key is \textit{(n,g)}
  \item The public (decryption) key is $(\gamma,\mu)$
\end{enumerate}

\subsection{Get User Roles Function}
This function is responsible for assigning the user's role and activate the assigned role for which the user belongs. If user belongs to any group role then validate the group authentication and assign to user. 
 \begin{figure}[H]
 \vspace{-1.5em}
    \begin{minipage}{0.5\textwidth}
      \includegraphics[width=\linewidth]{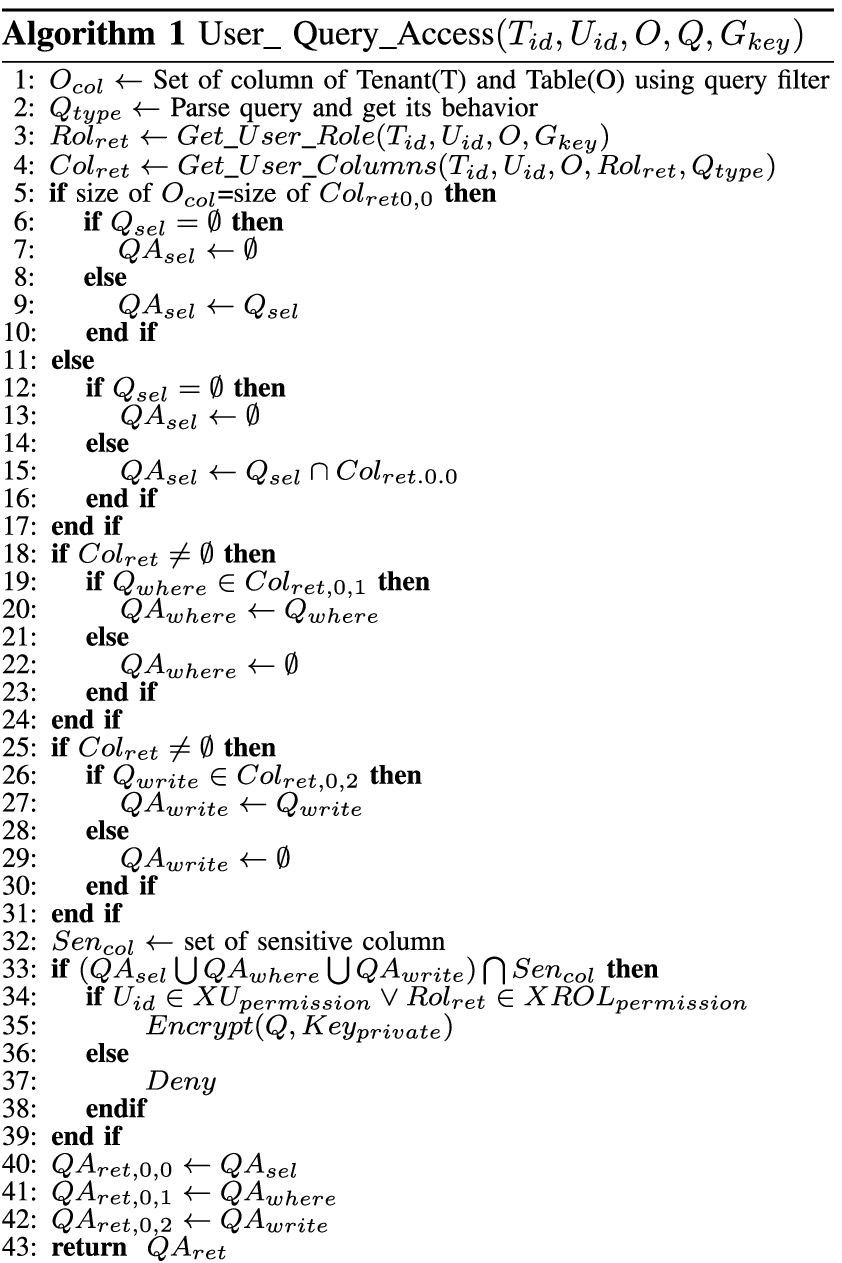}
    \end{minipage}
    \hfill
    \begin{minipage}{0.5\textwidth}
      \includegraphics[width=55mm]{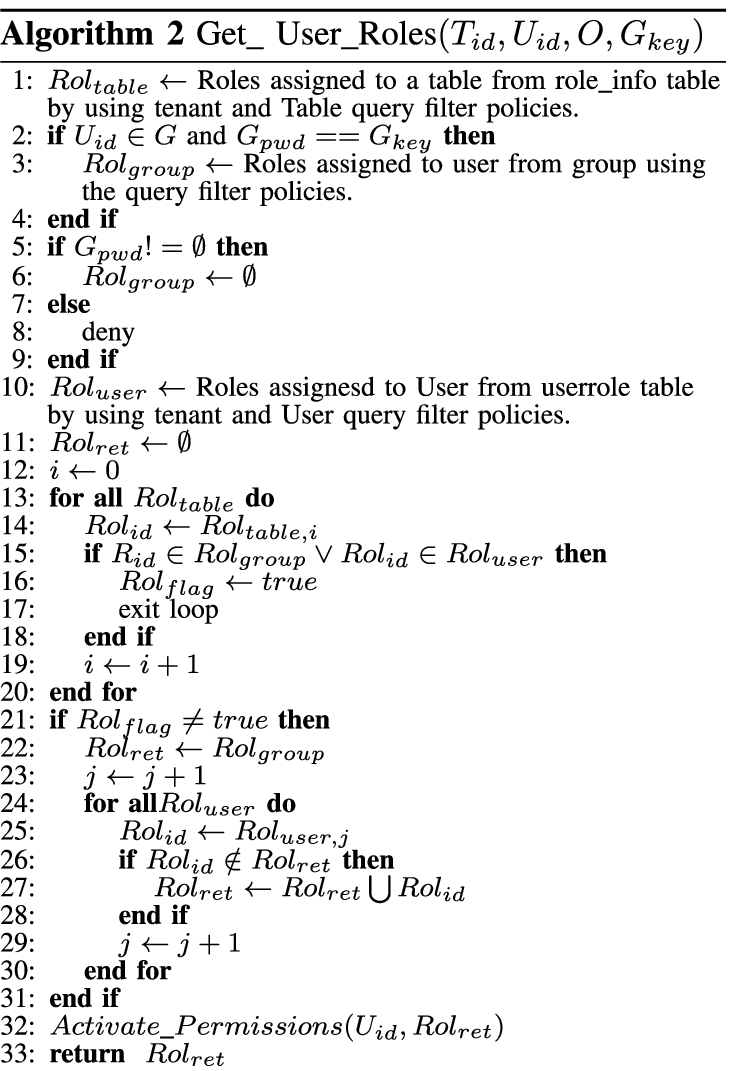}
      \fontsize{7}{7}\selectfont
      \vspace{12mm}
    \end{minipage}
    \vspace{-1.2em}
  \end{figure}
  
\subsection{Activate Permission and Deactivate permission}
The activation and deactivation functions are based on \cite{wu2013acaas}. The user's role access permission is included in proposed work. Usually, the user is assigned with more than one role, to provide the query access to the user so that the least privilege is met to complete the task to the user. The Nested Set model \cite{kamfonas1992recursive} with the use of tree data structure is used for maintaining the role hierarchy in the access control mechanism. The main advantage of using this model is that it requires a single entry for each role to manage the role hierarchy. The security policies enforce the role hierarchy like a regular role that have the least privilege and least access to the database. 

\par If a user $U_{id}$ has an immediate senior role in role hierarchy, then that role is assigned to role $Rol_{a}$ which needs to be activated, and an empty permission set is returned because the immediate senior role has default permission by policies and constraints to access database (like admin role). If the user  $U_{id}$ does not have immediate senior then for each junior and siblings the permission sets are calculated to activate the $Rol_{a}$. If multiple queries with the same roles are processed then the roles overlap and the deactivation function will not affect the other queries having the same roles.
 \begin{figure}
 \vspace{-1.5em}
    \begin{minipage}{0.5\textwidth}
      \includegraphics[width=\linewidth]{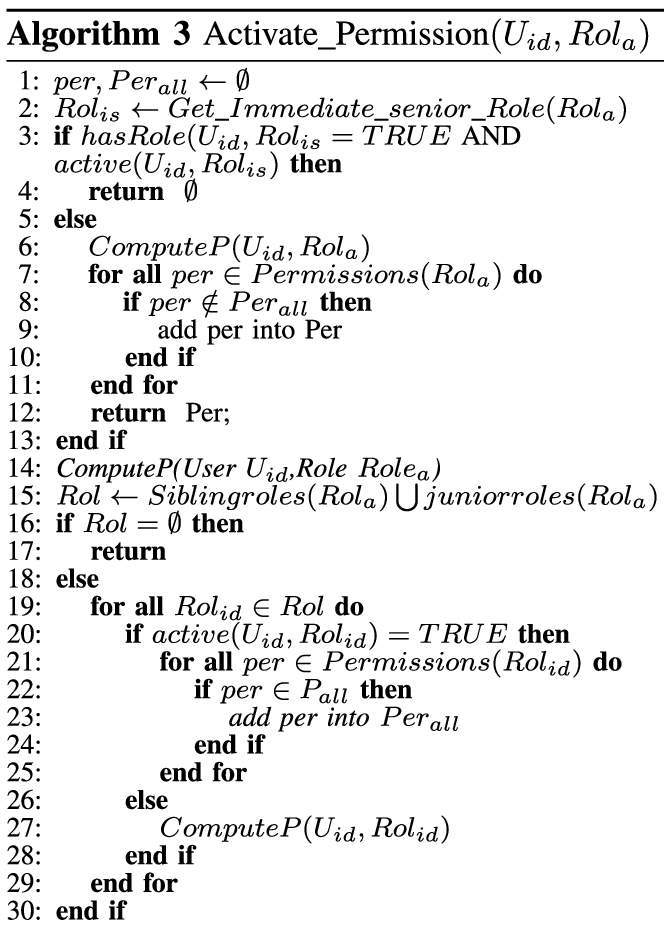}
    \end{minipage}
    \hfill
    \begin{minipage}{0.5\textwidth}
      \includegraphics[width=\linewidth]{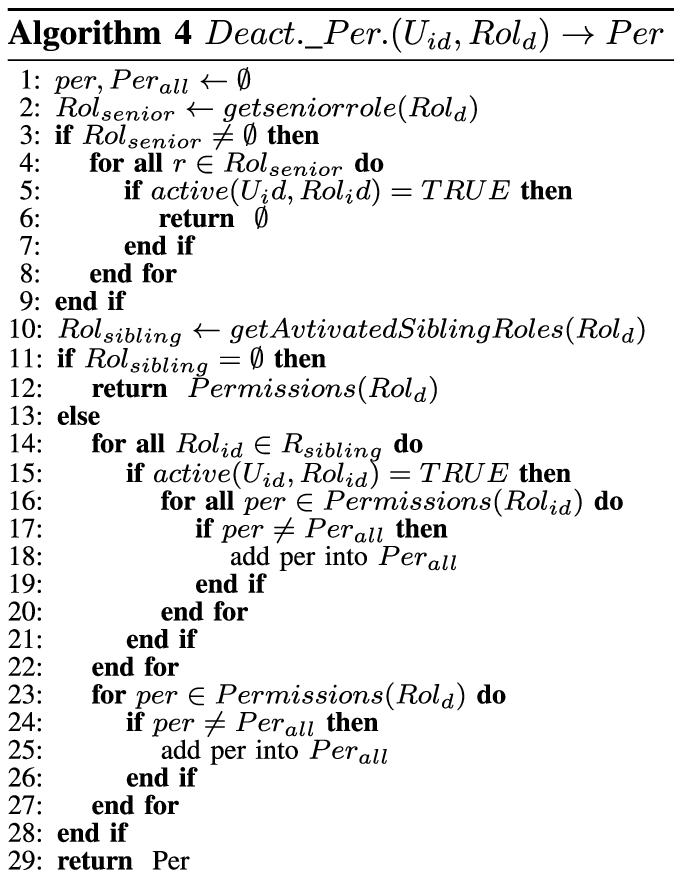}
      \vspace{3.5mm}
    \end{minipage}
  \vspace{-1.5em}
  \end{figure}

\par The role deactivation algorithm is shown in Algorithm 4.and during deactivation an empty set of permission is returned when any senior role to a role $Rol_d$, which required to deactivate for the user. Otherwise, if any role is needed to deactivate. For which the user that do not have any activated siblings roles, and then $Per_{all}$ is permission set containing all permission that has associated with activated sibling's role for the user. Moreover, with $Rol_d$ each permission is associated, if it does not belong to $Per_{all}$ then it adds to returned permission set.  

\subsection{Get User Columns}
The get user columns function is responsible for returning the set of accessible list of columns ($Col_{ret}$) which exist in different types of clauses (like select and where) in user's query.

\begin{figure}
    \vspace{-1.5em}
    \includegraphics[width=75mm]{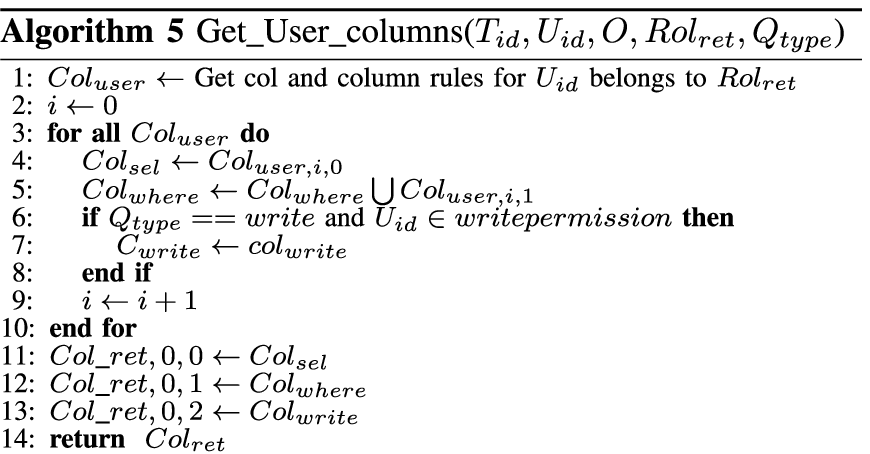}
    \vspace{-3em}
\end{figure}

\section{Implementation and Analysis}
A client-server architecture is considered for the implementation of proposed work. The server is a centralized one, but the database storage nodes are distributed by default distribution of MYSQL-Cluster database. The implementation of RBAC work is at the server side. The cryptographic access control is created using python, C++ and Crypto library.
\par The access control is a secure authentication mechanism that generates the session key using SHA-2 to encrypt the query at the client system. The session key is valid until the completion of the transaction. The server also encrypts the result (after successful execution of the query) using the same session key. Hence query and the result are secured during the communication. The proposed approach performs the group validation using the group key stored at the server. If the user belongs to the group, it has to provide the group key with the query. The group validation reduces the privilege escalation attacks because if the user escalate the role, and if that role belongs to any group, without group key the access control will deny the query. 
\par The proposed work manages the role hierarchy using the Nested set model. The attacks related to privileges is reduced due to the role hierarchy management with least privilege grant by adding minimum permissions for query execution. The activation and deactivation functions play a major role to protect from the escalation attack because by default roles are considered as deactivated roles and both functions need a user id to activate the role. The Paillier encryption algorithm is used to encrypt the sensitive data. So the sensitive data is secured during the execution of the query as well as in a database. Since the Paillier encryption has the properties of homomorphic encryption and works on the public-key cryptosystem, it protects the sensitive data from the cryptographic attacks such as chosen-plaintext attacks (IND-CPA) and adaptive chosen-cipher text attacks (IND-CCA2).
  \begin{figure}[!htb]
    \begin{minipage}{0.5\textwidth}
      \includegraphics[width=65mm]{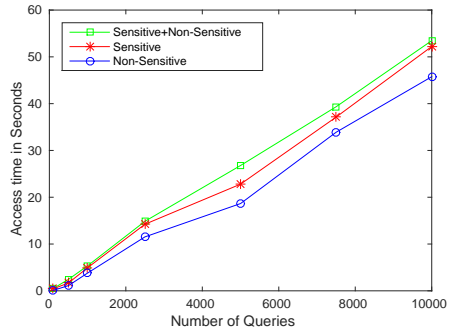}
      \caption{Access Time Vs Number of query}\label{fig:x1}
      \vspace{3mm}
    \end{minipage}
    \hfill
    \begin{minipage}{0.5\textwidth}
      \includegraphics[width=65mm]{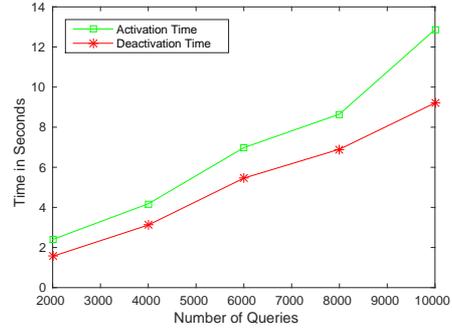}
      \caption{Activation and Deactivation Time Vs Number of Query}\label{fig:x2}
    \end{minipage}
  \end{figure}

\par Experiments are conducted using twenty-five different roles and eight groups are created for the tenant's database user. For the analysis, three type of queries are tested i) sensitive and non-sensitive data in query ii) only sensitive data in query iii) non-sensitive data in the query. The database has the four sensitive data columns. Fig. 3 shows number of queries vs. query access time and it is observed that there is little difference between query access time of sensitive data and non-sensitive data. This difference is due to extra computation time needed for sensitive data encryption. However this overhead provides the data confidentiality for the database owner. The role activation and deactivation time is represented in fig. 4. The role activation function and deactivation is responsible for providing integrity feature for proposed work. The role activation time is calculated by considering a maximum of 8 roles of the users and 3 groups of users. The queries contain sensitive attributes which need some extra permission that is required for the role activation. 

\section{Conclusion}

In this paper, a role-based access control for the DBaaS with data encryption using Paillier (partial homomorphic) public-key encryption is proposed. The approach is secure from the privilege escalation because of the role hierarchy management with least privilege grants. It protects from the SQL injection attacks since the role is to be activated to perform any query on database. The proposed approach works for the application as a service to provide secure query access control. The session key is used for securing the user query and on another side the group keys are used for checking whether the user belongs to groups. The role activation and deactivation functions help to maintain the role hierarchy. It adds the permission in the permission set. So, to perform any query access it adds minimum permission to execute the query and provide the least privilege for the task. At the server end data stored is in encrypted form. The partial homomorphic encryption used to provide confidentiality for sensitive data, encryption is secured from the IND-CPA and IND-CCA2. The limitation of work is that the user has to remember two or more keys.

\bibliographystyle{splncs03}
\bibliography{ref}

\begin{thebibliography}{10}
\providecommand{\url}[1]{\texttt{#1}}
\providecommand{\urlprefix}{URL }

\bibitem{bertino2011access}
Bertino, E., Ghinita, G., Kamra, A.: Access control for databases: concepts and
  systems, vol.~8. Now Publishers Inc (2011)

\bibitem{bertino2005database}
Bertino, E., Sandhu, R.: Database security-concepts, approaches, and
  challenges. Dependable and Secure Computing, IEEE Transactions on  2(1),
  2--19 (2005)

\bibitem{brodersen2004database}
Brodersen, K., Rothwein, T.M., Malden, M.S., Chen, M.J., Annadata, A.: Database
  access method and system for user role defined access (May~4 2004), uS Patent
  6,732,100

\bibitem{calero2010toward}
Calero, J.M.A., Edwards, N., Kirschnick, J., Wilcock, L., Wray, M.: Toward a
  multi-tenancy authorization system for cloud services. IEEE Security \&
  Privacy  8(6),  48--55 (2010)

\bibitem{echeverria2010permission}
Echeverria, V., Liebrock, L.M., Shin, D.: Permission management system:
  Permission as a service in cloud computing. In: Proceedings of the 2010 IEEE
  34th Annual Computer Software and Applications Conference Workshops. pp.
  371--375. IEEE Computer Society (2010)

\bibitem{gentry2009fully}
Gentry, C.: A fully homomorphic encryption scheme. Ph.D. thesis, Stanford
  University (2009)

\bibitem{hu2009towards}
Hu, L., Ying, S., Jia, X., Zhao, K.: Towards an approach of semantic access
  control for cloud computing. In: Cloud Computing, pp. 145--156. Springer
  (2009)

\bibitem{jia2012privacy}
Jia, Z., Pang, L., Luo, S.s., Zhang, J.y., Xin, Y.: A privacy-preserving access
  control protocol for database as a service. In: Computer Science \& Service
  System (CSSS), 2012 International Conference on. pp. 849--854. IEEE (2012)

\bibitem{kamfonas1992recursive}
Kamfonas, M.J.: Recursive hierarchies: The relational taboo. The Relational
  Journal  27 (1992)

\bibitem{matsumoto1998mersenne}
Matsumoto, M., Nishimura, T.: Mersenne twister: a 623-dimensionally
  equidistributed uniform pseudo-random number generator. ACM Transactions on
  Modeling and Computer Simulation (TOMACS)  8(1),  3--30 (1998)

\bibitem{takabi2010security}
Takabi, H., Joshi, J.B., Ahn, G.J.: Security and privacy challenges in cloud
  computing environments. IEEE Security and Privacy  8(6),  24--31 (2010)

\bibitem{wu2013acaas}
Wu, R., Zhang, X., Ahn, G.J., Sharifi, H., Xie, H.: Acaas: Access control as a
  service for iaas cloud. In: Social Computing (SocialCom), 2013 International
  Conference on. pp. 423--428. IEEE (2013)

\bibitem{yaish2013multi}
Yaish, H., Goyal, M.: Multi-tenant database access control. In: Computational
  Science and Engineering (CSE), 2013 IEEE 16th International Conference on.
  pp. 870--877. IEEE (2013)

\bibitem{yaish2011elastic}
Yaish, H., Goyal, M., Feuerlicht, G.: An elastic multi-tenant database schema
  for software as a service. In: IEEE ninth international conference on
  Dependable, autonomic and secure computing (dasc), 2011. pp. 737--743. IEEE
  (2011)

\end{thebibliography}

\end{document}